\documentstyle[amscd,amssymb,12pt]{amsart}
\input xypic \xyoption{curve}
\input epsf

\def\hcorrection#1{\advance\hoffset by #1 }
\def\vcorrection#1{\advance\voffset by #1 }

\vcorrection{-.7in}
\hcorrection{-.5in}

\theoremstyle{plain}
\newtheorem{th}{Theorem}[section]

\newtheorem{prop}{Proposition}[section]

\theoremstyle{definition}
\newtheorem{defin}{Definition}[section]

\theoremstyle{definition}

\theoremstyle{remark}

\numberwithin{equation}{section}

\begin{document}

\pagestyle{plain}
\addtolength{\footskip}{.3in}

\title{A Hopf algebra deformation approach to renormalization}
\author{Lucian M. Ionescu}
\author{Michael Marsalli}
\address{Department of Mathematics, Illinois State University, IL 61790-4520}
\email{lmiones@@ilstu.edu,marsalli@@math.ilstu.edu}


\begin{abstract}
We investigate the relation between 
Connes-Kreimer Hopf algebra approach to renomalization 
and deformation quantization.
Both approaches rely on factorization,
the correspondence being established at the level of Wiener-Hopf algebras,
and double Lie algebras/Lie bialgebras, via r-matrices.

It is suggested that the QFTs obtained via 
deformation quantization and renormalization 
correspond to each other in the sense of Kontsevich/Cattaneo-Felder \cite{Kon1,Cat1}.
\end{abstract}

\maketitle
\tableofcontents


\section{Introduction}\label{S:intro}
Connes-Kreimer's Hopf algebra approach to renormalization \cite{CKI,Kpqft,Kfg} 
allows to identify an ``algebraic interface'' to renormalization,
opening an avenue for connections with deformation quantization.

Renormalization is a factorization at the level of the 
convolution algebra of ``commutative points'' of the Hopf algebra of Feynman graphs, 
corresponding to a given renormalization scheme.
This Birkhoff decomposition / Wiener-Hopf factorization 
is an algebraic geometry analog of the infinitesimal factorization \cite{Sem} 
leading to quantization (for the global version, see Poisson-Lie groups and their factorization,
\cite{CP}, p.67).

Indeed,
the infinitesimal factorization is essentially equivalent to a Lie bialgebra structure,
and hence, via a ``standard procedure'', to quantization (\cite{EK,R}; \cite{CP}):
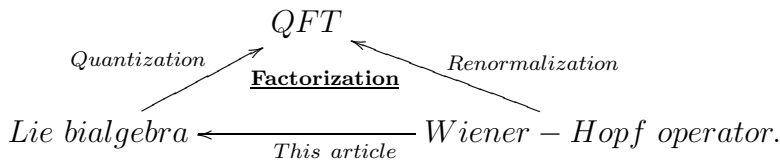
$$\diagram
& QFT & \\
Lie\ bialgebra \urto^{Quantization\quad}  & \ar@{}[u]|{{\bf \quad \underline{Factorization}}}
& \llto^{This\ article\quad} \ulto_{\qquad Renormalization} Wiener-Hopf\ operator.
\enddiagram$$
A prescription scheme satisfying the multiplicativity condition is nothing else but
a Baxter operator \cite{Rota},
leading to a solution of the modified classical Yang-Baxter equation,
and therefore to a coboundary Lie bialgebra structure and 
deformation quantization \cite{CP}, p.187.

The main result, the Factorization Theorem \ref{T:FT},
recognizes the factorization of Poisson-Lie groups (\cite{CP}, p.67)
corresponding to the Birkhoff decomposition associated to 
a renormalization prescription scheme.
The ``deformed antipode'', or rather inverse of a character (\cite{Kpqft}, p.19),
is interpreted as part of the ``decomposition'' of the convolution product.
This prompts for an interpretation of the multiplicativity condition: 
a classical Yang-Baxter equation in disguise.

The results presented here are expected to have an impact on the understanding of perturbative QFT, 
besides the obvious message that the details of the various receipts, 
computational schemes etc. of renormalization
are essentially irrelevant to the overall physics scheme.
It is expected that a unifying point of view including 
renormalization, deformation quantization and TQFTs will emerge.
The central notion in producing QFTs (as pictured above) is therefore 
the concept of {\em factorization} (algebraic/geometric, global/infinitesimal etc.).

The paper is organized as follows.

The algebraic interface to renormalization is recalled in \S\ref{S:ai},
following \cite{CKI,Kpqft}.

The factorization point of view is studied in \S\ref{S:dq},
where Wiener-Hopf operators are defined as a special case of Baxter operators.
Some of their properties are stated in Theorem \ref{T:WH}.
An infinitesimal factorization occurs \cite{Sem},
with the associated ``Hilbert transform'' satisfying 
a ``Poincare-Bertrand'' equation.
The Lie algebra correspondent is a modified Yang-Baxter equation.

This leads to a ``Factorization Theorem'' \ref{T:FT} on the convolution algebra,
as an algebraic source for the Birkhoff decomposition in the analytic framework.
The benefit is a prediction of a classical R-matrix,
opening the path to Lie bialgebra quantization.

We conclude in \S\ref{S:cd} with additional comments
on the relation between deformation quantization and renormalization.

\vspace{.2in}
{\bf Acknowledgments.}

L.I. would like to express our gratitude 
for the excellent research conditions at I.H.E.S.
where the present article was started after 
stimulating discussions with Kobi Kremnizer.

\section{An algebraic interface to renormalization}\label{S:ai}
We will start with a brief recall on renormalization following \cite{CKI,Kpqft},
using freely the concepts introduced in \cite{Ipqft}.

The main ``characters in the renormalization play'' are:
a Lie algebra $g$ generated by some Feynman graphs,
with the corresponding Hopf algebra $H=U(g^*)$, 
a set of {\em states on Feynman graphs}
(e.g. external structures, ``colors'' to distinguish various propagators etc.),
allowing to define characters $\phi:H\to A$ into a commutative algebra 
as {\em bare values of the Feynman integral} 
corresponding to {\em Feynman rules} for a given regularization scheme,
and a k-linear map $R:A\to A$ 
representing the {\em renormalization prescription} 
(e.g. minimal subtraction (MS) with zero subtraction point).

Then,
Connes and Kreimer proved in \cite{CKI} that 
the renormalized value of the Feynman integral $\phi_+$
is a factor of the Birkhoff decomposition:
$$\phi=(\phi^-)^{-1}\star\phi^+$$
of the bare value of the dimensional regularization of the Feynman integral $\phi$.
It is obtained via the convolution with the counter term:
$$\phi^+=S_R\star\phi.$$
Recall that,
the Birkhoff decomposition of a meromorphic function (relative to a disk)
is a factorization:
\begin{equation}\label{E:bd}
\phi_+=\phi\star \phi_-,
\end{equation}
with $\phi_\pm$ holomorphic inside/outside the disk,
agreeing on the common boundary.
More generally it can be defined with respect to a decomposition of an algebra
into subalgebras.
\begin{defin} $\phi:\phi^-\to \phi^+$ denotes the {\em Birkhoff decomposition} of
the character $\phi:H\to A$ relative to the decomposition $A=A_-\oplus A_+$
corresponding to the projectors $(R^+,R^-)$
iff $\phi^\pm$ are characters satisfying
the following relation:
\begin{equation}\label{E:genbd}
\phi_+=\phi\star\phi_-, \quad R^\pm\phi_\pm=\phi_\pm.
\end{equation}
\end{defin}
The renormalization procedure based on the minimal subtraction prescription
is essentially a Birkhoff decomposition,
with the ``finite part'' of the Feynman amplitude (without poles in the analytic renormalization picture)
provided by $\phi_+$.

The above counter term is a deformation of the inverse 
$\phi^{-1}=\phi\circ S$ (with respect to convolution),
and can be obtained inductively by the formula:
\begin{equation}\label{E:CKdef}
S^\phi_R(\Gamma)=R(-\phi(\Gamma)-\sum_{\gamma\hookrightarrow\Gamma\to\gamma'} S^\phi_R(\gamma)\phi(\gamma')).
\end{equation}
In the above formula, 
the sum runs over certain decompositions of the graph $\Gamma$.
These possible decompositions can be encoded in a coproduct $\Delta$ 
(\cite{Kpqft}; see also \cite{Ipqft}).

Now the linear operator $R$ has to satisfy a multiplicative constraint:
\begin{equation}\label{E:baxter}
R(x)R(y)+R(xy)=R(R(x)y)+R(xR(y)),\quad x,y\in A,
\end{equation}
which ensure that $S^\phi_R$ is again a character (\cite{Kchen}, p.14).
This turns $(A,R)$ into a Baxter algebra (\cite{Kfg}, p.18).
Moreover, $R$ is usually idempotent (e.g. MS-prescription).

\section{Relation with deformation quantization}\label{S:dq}
Our point of view suggests to interpret the deformed inverse $S^\phi_R$
as part of the {\em factorization} enabling a Lie bialgebra deformation quantization,
following \cite{CP}, p.67.
In order to do this, rewrite Equation \ref{E:CKdef} as follows:
\begin{equation}\label{E:defantipode}
S^\phi_R(\Gamma)=R\circ m(-\phi(\Gamma)\otimes 1)
+\sum_{\gamma\hookrightarrow\Gamma\to\gamma'} R\circ m(S^\phi_R(\gamma)\otimes \phi(\gamma')).
\end{equation}
Now $\phi$ is a character of the convolution algebra $H^*=(Hom(H,A),\star,\eta)$, with:
$$<f\star g,X>_H=m_A\circ <f\otimes g,\Delta X>,\quad f,g\in Hom(H,A), X\in H,$$
or briefly $\star=m_A\circ (\cdot\otimes\cdot)\Delta$.
Here $H$ is regarded merely as a coalgebra and $A$ as an algebra with multiplication
denoted by $m_A$.

The multiplication $m_A$ can be truncated using $R$, yielding a new convolution:
$$m_+=R\circ m, \quad *_+=m_+\circ (\cdot\otimes\cdot)\Delta.$$
Although not associative when $R$ is a (nontrivial) projector as above,
the new multiplication $m_+$ is associative when 
$R$ is the ``Hilbert transform'' associated with a
Wiener-Hopf operator (see (iii) of Theorem \ref{T:WH}).
\begin{prop}\label{P:antipode}
The deformed inverse $S^\phi_R$ is the inverse relative to the
deformed convolution product $*_+$:
$$S^\phi_R=\phi^{-1}\quad in\ (H^*_+,*_+).$$
\end{prop}
In particular, in dimensional regularization,
a natural implementation of the requirement that 
`` ... it does not modify the pole terms in $\epsilon$'' \cite{Kfg}, p.18,
is to consider $R=R_-$ the projection on the pole part of the Taylor expansion.

The role of $R$ and the relation with the deformation picture 
will be investigated next.

%
%
\subsection{Wiener-Hopf operators}\label{S:WH}
We will address the structure of Wiener-Hopf operators
and the connection with Yang-Baxter equation.

Following \cite{Rota}, p.505,
we introduce the following definition.
\begin{defin}
An idempotent $R\in End(A)$ satisfying the Baxter identity is
called a {\em Wiener-Hopf operator}.
\end{defin}
Note that, citing from \cite{Rota}, 
``In fact it was the algebraic study of Wiener-Hopf operators that motivated
Glen Baxter to single out the Baxter identity."
Operators of this form arise in the theory of Wiener-Hopf factorization
(\cite{Gflu}, p.483),
which essentially corresponds to the Birkhoff decomposition via 
the Cayley transform.
 
Any Wiener-Hopf operator corresponds to a decomposition
of the algebra (\cite{Gflu}, p.181),
yielding a decomposition of the associated Lie algebra.

Since to quantize a Lie bialgebra structure is needed
(double Lie algebra \cite{Sem}) ,
this Wiener-Hopf structure seems to play in renormalization 
the role the double construction plays in quantization.
For details we send the reader to \cite{Sem}, p.263 
(Infinitesimal Form of the Factorization Theorem, r-matrices, 
Lie bialgebra structure and double Lie algebras),
including the relation with Lax equations, Riemann problem etc. (p.262).

The ``missing link'', to be studied next,
is the connection between the Baxter operator $R$
and the classical r-matrix in the sense of \cite{Sem}, p.260,
satisfying the modified Yang-Baxter equation (loc. cit. p.262):
\begin{equation}\label{E:MYBE}
[Rx,Ry]-R([Rx,y]+[x,Ry])=-[x,y]\tag{MYBE}.
\end{equation}
Any solution of MYBE determines a {\em modified bracket}:
\begin{equation}\label{E:mlb}
[x,y]_R=[Rx,y]+[x,Ry], \quad x,y\in A.
\end{equation}
satisfying the Jacoby identity (see \cite{Sem}, p.262).

%
%
\subsection{Associative algebra level}\label{S:aal}
The structure of a Wiener-Hopf algebra is equivalent 
to having a decomposition of the algebra as a sum of two subalgebras.

If $R$ is an idempotent of $End(A)$, 
consider the associated decomposition $A=A_-\oplus A_+$.
\begin{defin}
The symmetry about the ``positive part'' of the algebra $(A,m)$
under the decomposition along the projector $R_+=R$, 
will be called the associated {\em Hilbert transform}:
\begin{equation}\label{E:HT}
U=R_+-R_-.
\end{equation}
Here $R_-=I-R$ is the complementary projector.
\end{defin}
An example of such an idempotent $R_+$ is 
the projection which cuts the pole part of a meromorphic function.
\begin{prop}\label{T:WH}
Let $R$ be an idempotent of $A$,
determining the above direct sum decomposition as k-modules.
Then the following statements are equivalent:

(i) $R$ is a Wiener-Hopf operator, i.e. it satisfies the Baxter identity;

(ii) $A_\pm$ are subalgebras in $A$;

(iii) The Hilbert transform satisfies the Poincare-Bertrand 
identity (PBI - following \cite{Sem}, p.263):
\begin{equation}\label{E:PBI}
Rx\cdot Ry+xy=R(Rx\cdot y+x\cdot Ry), \quad x,y\in A.
\end{equation}
Moreover, 
in this case the following {\em twisted multiplication} is associative:
\begin{equation}\label{E:defmult}
m_U(x,y)=U(x)\cdot y+x\cdot U(y), \quad x,y\in A.
\end{equation}
\end{prop}
\begin{pf}
Since both equations, Baxter equation and PBI, are linear,
we can restrict to the corresponding subspaces,
and establish the statements using a case-by-case analysis.
For instance,
if $x\in A_+$ and $y\in A_\pm$,
then a direct computation shows that 
the Baxter identity is equivalent to $R(xy)=xy$, etc.

The case-by-case analysis ($x,y,z\in A_\pm$) to establish the last statement
may be reduced using the symmetry of the associator,
and noting that on $A_+$, $m_U=2m$, on $A_-$ $m_U=-2m$,
while on $A_+\otimes A_-$ $m_U$ vanishes
(see also \cite{Sem}, p.263, Proposition 6).
This completes the proof.
\end{pf}
Now we will turn our attention to the infinitesimal level.

%
%
\subsection{Lie algebra level}\label{S:lal}
Applying the alternation functor
$``Alt=\sum_{\sigma\in \Sigma}\epsilon(\sigma)'',$
and in particular passing from associative algebras to Lie algebras
($L:Alg<->Lie:U$),
we will obtain the corresponding Lie algebra picture.
\begin{th}\label{P:lie}
Let $R\in End(A)$ and $(g,[,])=Alt(A,m)$, 
i.e. $[x,y]=m(x,y)-m(y,x), x,y\in A$.
If $R$ is a Wiener-Hopf operator, then:

(i)$R$ satisfies the modified Yang-Baxter equation:
\begin{equation}
[Rx,Ry]+R[x,y]=R([,]_R), \quad x,y\in g,
\end{equation}
and the modified bracket $[,]_U$ is a Lie bracket.

(ii) $(g,[,]_R)$ is a double Lie algebra corresponding to the decomposition
$$g=g_-\oplus g_+, \quad g_\pm=Alt(A_\pm).$$
\end{th}
\begin{pf}
(i) is plainly obtained by alternating Baxter equation.

Regarding the ``infinitesimal factorization'' of part (ii),
we refer the reader to Proposition 5, \cite{Sem}, p.262.
\end{pf}
As a consequence, $R$ is a r-matrix,
providing an infinitesimal factorization $g=g_-\oplus g_+$
via the Lie-algebra homomorphisms $R\pm 1$ (\cite{Sem}, 263).
Moreover, note that an r-matrix exists iff 
the analogous Lie algebra factorization 
is uniquely solvable (loc. cit. p.264).

As proved in Proposition 3, \cite{Sem}, p.261,
double Lie algebras and Lie bialgebras structures are equivalent
in the case of skew symmetric solutions of MYBE,
in the presence of an invariant inner product.
This opens the avenue of deformation quantization,
towards a QFT ``a la Moyal''.

%
%
\subsection{The factorization property}
We return to the Birkhoff decomposition,
as interpreted in Proposition \ref{P:antipode}.

The truncated multiplication $m_+$ is not associative,
since $R$ is a projector satisfying the Baxter identity.
It is ``half'' of the factorization picture,
and by reasons of symmetry,
one has to consider the complementary projector $R_-=Id-R_+$
with the associated truncated multiplication $m_-=R_-\circ m$.

Define the two inversion maps relative to the corresponding convolutions:
$$S_\pm:H^*\to H^*,\quad S_\pm(\phi)=\phi^{-1}\ in\ H^*_\pm.$$
Note that since $R_++R_-=id$ (trivial overlapping of the two ``charts'')
the ``transition map'' $T=(S_-)^{-1}S_+$ is the identity map on $H^*$,
establishing the following ``Factorization Theorem'' 
(compare with the Poisson-Lie group factorization \cite{CP}, p.67).
%
\begin{th}\label{T:FT}
For any Wiener-Hopf algebra $(A,m,R)$ and coalgebra $(H,\Delta)$,
the convolution algebra $(H^*,\star)$ is {\em factorizable},
i.e. there is a canonical Birkhoff decomposition:
$$\phi=\phi^{-1}_-\star\phi_+,\quad \phi\in Hom(H,A),$$
where $\phi_\pm=S_\pm(\phi)$ are the inverses relative to the
truncated convolutions $*_\pm$.
\end{th}
Next, we will only ponder on some possible implications.

\section{Conclusions}\label{S:cd}
Among the notable consequences of the above factorization,
one expects the existence of a canonical classical R-matrix
similar to the one present in the context of Poisson-Lie groups
\cite{CP}, p.67.
This would allow to explore the deformation quantization path on 
the triangle diagram from the introduction.

In fact the two quantizations obtained via star-products and renormalization
were identified in the introduction for simplicity.
A ``resolution'' of the tip of the triangle is provided in \cite{Kon1,Cat1}:
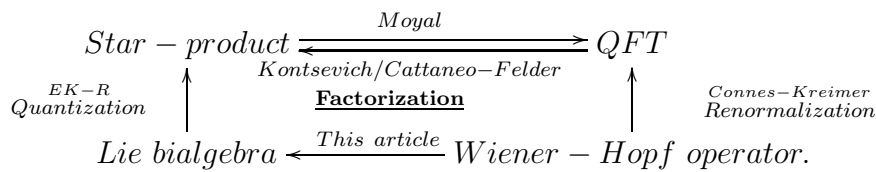
$$\diagram
Star-product \ar@<2pt>[rr]^{Moyal} &  & QFT \ar@<2pt>[ll]^{Kontsevich/Cattaneo-Felder} \\
Lie\ bialgebra \uto^{\overset{EK-R}{Quantization}\quad}
  & \ar@{}[u]|{{\bf \quad \underline{Factorization}}}
& \llto_{This\ article\qquad} \uto_{\qquad \overset{Connes-Kreimer}{Renormalization}} 
Wiener-Hopf\ operator.
\enddiagram$$
For details on the left edge (quantization) see \cite{CP}, Ch.6, p.170,
and references therein (\cite{EK,R} etc.).

The suggested ``commutativity'' of the above diagram needs 
a more detailed investigation,
clarifying the role of the classical R-matrix expected above.
This will support the claim that renormalization 
is independent on the chosen prescription $R$,
and contribute to the understanding of the renormalization group
in the context of Weyl's program in quantum physics \cite{M}.

Moreover, 
it was suggested in \cite{Ipqft} that the Feynman graphs (as a category of cobordism)
should be thought of as a (quantum) ``space-time''.
This interpretation is consistent with the duality ``spectrum - convolution algebra''
present in the deformation point of view of renormalization.
A more detailed account will appear elsewhere.




\begin{thebibliography}{33}

\bibitem[1]{CKI} A. Connes and D. Kreimer,
Renormalization in quantum field theory and the Riemann-Hilbert problem I,
hep-th/9912092; Comm. Math. Phys. 210 (2000), no. 1, 249--273.

\bibitem[2]{Kpqft} D. Kreimer,
Combinatorics of (perturbative) Quantum Field Theory, hep-th/0010059;
Renormalization group theory in the new millennium,
IV (Taxco, 1999). Phys. Rep. 363 (2002), no. 4-6, 387--424.

\bibitem[3]{Kfg} D. Kreimer,
Structures in Feynman Graphs - Hopf algebras and symmetries,
hep-th/0202110.

\bibitem[4]{Sem} M. A. Semenov-Tyan-Shanskii,
What is a classical R-matrix?, Functional Anal. Appl. 17 (1983), no. 4, 259--272.

\bibitem[5]{CP} V. Chari and A. Pressley,
A guide to quantum groups, Cambridge Univeristy Press, 1994.

\bibitem[6]{EK} P. Etingof and D. Kazhdan, 
Quantization of Lie bialgebras. I. Selecta Math. (N.S.) 2 (1996), no. 1, 1--41.

\bibitem[7]{R} N. Reshetikhin,
Quantization of Lie bialgebras, Internat. Math. Res. Notices 1992, no. 7, 143--151.

\bibitem[8]{Rota} Gian-Carlo Rota,
Baxter operators, and introduction, p.504-512,
in {\em Gian-Carlo Rota on combinatorics - Introductory papers and commentaries},
Joseph P.S. Kung editor,
1995, Birkhauser, Boston-Basel-Berlin.

\bibitem[9]{Ipqft} L. M. Ionescu,
Perturbative quantum field theory and integrals on configuration spaces,
hep-th/0307062.

\bibitem[10]{Kchen} D. Kreimer,
Chen's Iterated Integral represents the Operator Product Expansion,
hep-th/9901099; Adv. Theor. Math. Phys. 3 (1999), no. 3, 627--670.

\bibitem[11]{Gflu} Gian-Carlo Rota,
Fluctuation theory and Baxter algebras, p.481-503,
in {\em Gian-Carlo Rota on combinatorics - Introductory papers and commentaries},
Joseph P.S. Kung editor,
1995, Birkhauser, Boston-Basel-Berlin.

\bibitem[12]{Kon1} M. Kontsevich,
Deformation quantization of Poisson manifolds I,
q-alg/9709040.

\bibitem[13]{Cat1} Alberto S. Cattaneo and Giovanni Felder,
A path integral approach to the Kontsevich quantization formula,
math.QA/9902090; Comm. Math. Phys. 212 (2000), no. 3, 591--611.

\bibitem[14]{M} G. W. Mackey,
Weyl's program and modern physics,
Differential geometrical methods in theoretical physics (Como, 1987), 11--36, 
NATO Adv. Sci. Inst. Ser. C Math. Phys. Sci., 250, 
Kluwer Acad. Publ., Dordrecht, 1988. 

\end{thebibliography}
\end{document}